\documentclass[
    ,final            % use final for the camera ready runs
  ]
  {aipproc}

\layoutstyle{6x9}

\begin{document}

\title{Recent developments in quasi-elastic scattering 
around the Coulomb barrier}


\author{K. Hagino}{
address={Department of Physics, Tohoku University,
Sendai 980-8578, Japan}
}

\begin{abstract} 
We discuss two recent topics on 
heavy-ion quasi-elastic scattering at energies around the Coulomb barrier. 
The first topic is an application of quasi-elastic scattering at 
deep-subbarrier energies to extracting the surface diffuseness parameter 
of the nucleus-nucleus potential. 
The second topic is a coupled-channels analysis 
for the quasi-elastic barrier distribution 
for the $^{70}$Zn + $^{208}$Pb reaction. 
We show that the coupled-channels calculations which include the 
multi-phonon excitations in the colliding nuclei reproduce reasonably well the experimental 
excitation function for quasi-elastic scattering at backward angles 
and the barrier distribution for this reaction. 
\end{abstract}

\maketitle

%%%%%%%%%%%%%%%%%%%%%%%%%%%%%%%%%%%%%%%%%%%%
%% MAINMATTER
%%%%%%%%%%%%%%%%%%%%%%%%%%%%%%%%%%%%%%%%%%%%

\section{Introduction}

The internal structure of colliding nuclei strongly influences  
heavy-ion collisions at energies around the Coulomb barrier. 
A well known example is a reaction of a deformed nucleus. In this
case, the nucleus-nucleus potential depends on the orientation angle of the 
deformed nucleus with respect to the beam direction. 
Assuming that the orientation angle does not change during the 
collision, the cross section can then be obtained by averaging 
the contribution from all possible angles \cite{W73,ARN88,RHT01}. 
In this picture, 
the relative motion between the colliding nuclei experiences 
many distributed potential barriers depending on the orientation 
angle of the target nucleus, instead of a single barrier. 
To a good approximation, 
the concept of barrier 
distribution can be extended also to 
systems with a non-deformed target \cite{DHRS98,BT98,HB04,DLW83}, 
where the distribution originates from the coupling between the 
relative motion and 
several intrinsic degrees of freedom such as collective
inelastic excitations of the colliding nuclei and/or transfer
processes. 

In Ref.  \cite{RSS91}, Rowley, Satchler, and Stelson argued that 
a barrier distribution can be directly extracted 
from a measured fusion excitation function 
$\sigma_{\rm fus}(E)$, by taking the second derivative of the product
$E\sigma_{\rm fus}(E)$ with respect to the center-of-mass energy $E$, that is, 
$d^2(E\sigma_{\rm fus})/dE^2$. This has stimulated many high precision 
measurements of fusion cross section, so that the second derivative 
is meaningful \cite{DHRS98,L95}. 
The extracted barrier distributions have revealed that 
the concept indeed holds and the barrier distribution 
provides a nice tool to investigate the fusion dynamics of the 
entrance channel. 
It was also shown recently that the concept of barrier distribution 
still retains even in massive systems, such as $^{100}$Mo +
$^{100}$Mo \cite{RGH06}. 

A similar barrier distribution can be extracted also using the 
quasi-elastic scattering \cite{TLD95,HR04}. 
The quasi-elastic scattering is 
a sum of elastic, inelastic, transfer, and breakup processes, and 
is a good counterpart 
of heavy-ion fusion reaction \cite{ARN88}. 
A major difference is that the
quasi-elastic scattering is related to the reflection probability 
of the Coulomb barrier, while the fusion is related to the 
transmission. Since the penetration and reflection probabilities are 
related to each other due to the flux conservation, similar
information can be obtained both from fusion and quasi-elastic 
scattering. 

In this contribution, we 
discuss two recent theoretical 
activities on heavy-ion quasi-elastic 
scattering at sub-barrier energies. 
We first present our recent systematic analyses on heavy-ion 
quasi-elastic scattering at deep-subbarrier energies, 
in aiming at extracting the surface diffuseness parameter 
of inter-nuclear potential \cite{WHD06,hagino05}. 
We then discuss coupled-channels calculations for 
the $^{70}$Zn + $^{208}$Pb reaction, for which the quasi-elastic 
barrier distribution has recently been obtained experimentally \cite{I06}. 

\section{Quasi-elastic barrier distributions}

Before we proceed, let us first summarize the 
theoretical formulas for quasi-elastic barrier distribution. 
In the eigenchannel representation of the coupled-channels 
method, the fusion and quasi-elastic cross sections are given 
as a weighted sum of the cross sections 
for uncoupled eigenchannels \cite{ARN88,DHRS98,BT98,HB04,DLW83}.  
That is, 
\begin{eqnarray}
\sigma_{\rm fus}(E)&=&\sum_\alpha w_\alpha 
\sigma_{\rm fus}^{(\alpha)}(E), \label{crossfus}\\
\sigma_{\rm qel}(E,\theta)&=&\sum_\alpha w_\alpha 
\sigma_{\rm el}^{(\alpha)}(E,\theta), \label{crossqel}
\end{eqnarray}
where $\sigma_{\rm fus}^{(\alpha)}(E)$ and 
$\sigma_{\rm el}^{(\alpha)}(E,\theta)$
are the fusion and the elastic cross sections for a potential 
in the eigenchannel $\alpha$. 
Notice that the same weight factors $w_\alpha$ appear both in 
Eqs. (\ref{crossfus}) and (\ref{crossqel}). 
This is a generalization of well-known orientation average formula 
for a system with deformed target, 
\begin{equation}
\sigma(E)=\int^1_0d(\cos\theta_T)\sigma(E;\theta_T),
\end{equation}
where $\theta_T$ is the orientation of the deformed target and 
represents a continuous variable for $\alpha$ in Eqs. (\ref{crossfus})
and (\ref{crossqel}). 

The idea of barrier distribution is led by the fact that the
classical cross sections for fusion and quasi-elastic scattering 
for a single potential barrier 
are given by 
\begin{equation}
\sigma^{cl(0)}_{\rm fus}(E)=\pi
R_b^2\left(1-\frac{V_b}{E}\right)\,\theta(E-V_b),
\end{equation}
and 
\begin{equation}
\sigma_{\rm el}^{cl(0)}(E,\pi)=\sigma_R(E,\pi)\,\theta(V_b-E), 
\label{el}
\end{equation}
respectively\cite{HR04}. Here, 
$R_b$ and $V_b$ are the barrier position and the barrier height for 
the $s$-wave scattering (thus the scattering angle is set to be $\pi$
in Eq. (\ref{el})), respectively, and 
$\sigma_R(E,\pi)$ is the Rutherford cross section. 
These yield \cite{RSS91,TLD95}, 
\begin{eqnarray}
D_{\rm fus}(E)&\equiv& \frac{d^2}{dE^2}[E\sigma_{\rm fus}(E)]
=\sum_\alpha w_\alpha \pi
\left[R^{(\alpha)}_b\right]^2\,\delta(E-V_b^{(\alpha)}),  \label{Dfus}\\
D_{\rm qel}(E)&\equiv& 
-\frac{d}{dE}\left(\frac{\sigma_{\rm el}(E,\pi)}{\sigma_R(E,\pi)}\right)
=\sum_\alpha w_\alpha\,\delta(E-V_b^{(\alpha)}). 
\label{Dqel}
\end{eqnarray}
Evidently, these functions provide information on how potential 
barrier heights are distributed, and 
are called fusion and quasi-elastic barrier distributions, 
respectively. In realistic situations, 
the quantum (tunneling) effect smears the delta function in 
Eqs. (\ref{Dfus}) and (\ref{Dqel}). 
Moreover, the effect of nuclear potential has to be taken 
into account in 
quasi-elastic cross sections in Eq. (\ref{Dqel}) \cite{HR04}. 
Nevertheless, from the derivation, it is apparent that the fusion 
and quasi-elastic barrier distributions behave in a similar way. 
This is demonstrated in Fig. 5 in Ref. \cite{HR04} 
for the $^{16}$O + $^{154}$Sm system.  

In actual experiments, it is impossible to put a detector at a 
scattering angle $\pi$. One can, however, 
scale a cross section in energy by taking 
into account the centrifugal correction. 
Estimating the centrifugal potential at the distance of closest 
approach for the Rutherford scattering, $r_c$, 
the effective energy may be expressed as \cite{TLD95}
\begin{equation}
E_{\rm eff}\sim E
-\frac{\lambda_c^2\hbar^2}{2\mu r_c^2} 
=2E\frac{\sin(\theta/2)}{1+\sin(\theta/2)}.
\label{Eeff}
\end{equation}
Therefore, one expects that the function 
$-d/dE (\sigma_{\rm el}/\sigma_R)$ 
evaluated 
at an angle $\theta$ will correspond to the quasi-elastic barrier 
distribution 
at the effective energy given by eq. (\ref{Eeff}). 

%\begin{figure}
%\includegraphics[scale=0.4,clip]{fig1}
%\caption{
%(a) The fusion barrier distribution 
%for the $^{16}$O + $^{154}$Sm reaction, 
%obtained with the
%orientation-integrated formula with $\beta_2=0.306$ and $\beta_4$=
%0.05. 
%(b) Same as Fig. 1(a), but for the quasi-elastic barrier
%    distribution.  
%%Experimental data are from Ref. {\protect\cite{TLD95}}. 
%(c) Comparison between 
%the barrier distribution for fusion (solid line)
%    and that for quasi-elastic scattering (dashed line). 
%These functions are both normalized to unit area in the energy interval 
%between 50 and 70 MeV.}
%\end{figure}

\section{Inter-nucleus potential and deep-subbarrier quasi-elastic 
scattering}

Let us now discuss 
the application of deep-subbarrier quasi-elastic scattering to 
the problem of surface 
diffuseness anomaly in heavy-ion potential \cite{HDGHMN01}.
For calculations of 
elastic and inelastic scattering,
which are sensitive only to the surface region of 
the nuclear potential,  
the diffuseness parameter 
of around 0.63 fm has been conventionally employed 
\cite{broglia91,Christensen76}.
This value of surface diffuseness parameter has been well accepted, 
partly because it is consistent with 
a double folding potential \cite{SL79}. 
In contrast, 
a recent systematic study has shown that 
experimental data for heavy-ion fusion reactions 
require a much larger value of the diffuseness parameter, 
ranging between 0.75 and 1.5 fm, 
as long as the Woods-Saxon 
parameterization is employed \cite{newton04}.

Since quasi-elastic scattering and fusion are complementary to each 
other, it is of interest to investigate this problem 
using quasi-elastic scattering. 
In doing so, we are particularly interested in the deep 
sub-barrier region  \cite{WHD06,hagino05}. 
At these energies, 
the cross sections of (quasi-)elastic scattering are close to the 
Rutherford cross sections, with small deviations 
caused by the effect of nuclear interaction, $V_N$. 
This effect can be taken into account 
by the semiclassical perturbation theory \cite{HR04,LW81}, which leads
to 
\begin{equation}
\frac{d\sigma_{\rm el}(E,\theta)}{d\sigma_R(E,\theta)}
\sim
1+\frac{V_N(r_c)}{ka}\,
\frac{\sqrt{2a\pi k\eta}}{E},
\end{equation}
where 
$k=\sqrt{2\mu E}/\hbar$, $\mu$ being the reduced mass. 
$\eta$ is the Sommerfeld parameter, and $r_c$ is the distance of
closest approach. 
This formula shows that 
the deviation of the elastic cross sections
from the Rutherford ones is 
sensitive predominantly to the surface region 
of the nuclear potential, especially to the surface diffuseness 
parameter $a$. 
There is another advantage of using the deep sub-barrier 
data. 
That is, 
the effect of channel coupling on 
quasi-elastic scattering can be disregarded 
at these energies, since the reflection 
probability is almost unity irrespective of 
the presence of channel couplings \cite{hagino05}. 
From these considerations, it is evident that 
the effect of surface diffuseness parameter can be studied 
in a transparent and unambiguous way using the large-angle quasi-elastic
scattering at deep sub-barrier energies. 

\begin{figure}
  \includegraphics[height=.45\textheight]{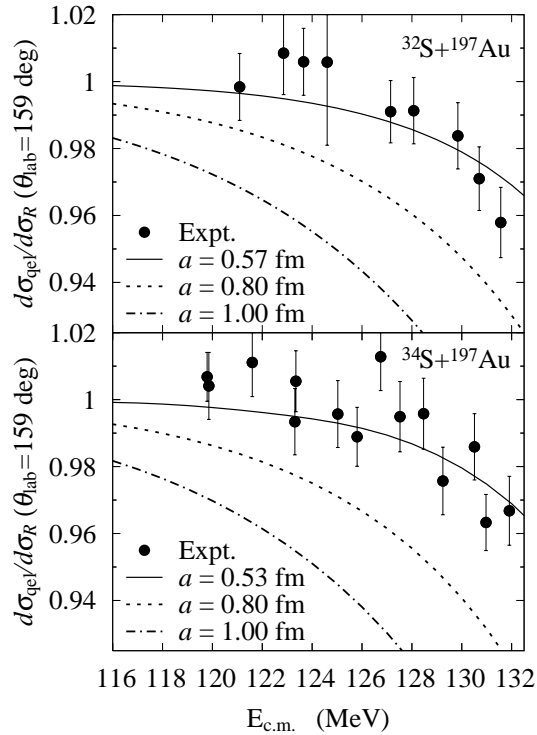}
  \caption{
The ratio of the quasi-elastic to the Rutherford cross sections
at $\theta_{\rm lab}=159^\circ$  
for the $^{32}$S + $^{197}$Au (the upper panel) reaction
and for the $^{34}$S + $^{197}$Au (the lower panel) reaction.
%The experimental data are taken from Ref. \cite{schuck02}. 
%The solid line results from using a diffuseness parameter obtained by 
%performing a least-square fit to the data.
%The dotted and the dot-dashed lines are obtained with 
%the diffuseness parameter of $a$ = 0.80 fm and $a$ = 1.00 fm,
%respectively. 
}
\end{figure}

Figure 1 compares the experimental data with
the calculated cross sections obtained with 
different values of the surface diffuseness 
parameter 
for the $^{32}$S + $^{197}$Au system (the upper panel) 
and the $^{34}$S + $^{197}$Au system (the lower panel). 
In order to analyze the experimental data 
at deep sub-barrier energies,
we use a one-dimensional optical potential 
with the Woods-Saxon form. 
Absorption following transmission through the barrier
is simulated by an imaginary potential that is 
well localized inside the Coulomb barrier.
The best fitted values for the surface diffuseness parameter 
are $a=0.57\pm 0.04$ fm and $a=0.53\pm 0.03$ fm 
for the $^{32}$S and $^{34}$S + $^{197}$Au reactions, respectively. 
The cross sections obtained with these surface diffuseness 
parameters are denoted by the solid line in the figure. 
The dotted and the dot-dashed lines are calculated with
the diffuseness parameter of $a$ = 0.80 fm and $a$ = 1.00 fm,
respectively.
It is clear from the figure that 
these spherical systems 
favor the standard value of the surface diffuseness parameter,
around $a=$ 0.60 fm. 
The calculations with the larger diffuseness 
parameters underestimate the quasi-elastic cross sections 
and are not consistent with the energy dependence of the 
experimental data.
We obtain a similar conclusion for the 
$^{32,34}$S + $^{208}$Pb and 
$^{16}$O + $^{208}$Pb systems\cite{WHD06}. 
This indicates that the double folding procedure is valid at least 
in the surface region and for spherical systems which we studied. 

For deformed systems, such as 
$^{16}$O + $^{154}$Sm, $^{186}$W, on the other hand, we found that 
the surface diffuseness 
parameter of $a=1.14 \pm 0.03$ fm and 
0.79 $\pm 0.04$ fm for the former and for the latter, respectively, 
is required in order to account for the experimental data 
\cite{WHD06,hagino05}. 
Although these large values of surface diffuseness parameter are 
consistent with that extracted from 
fusion, the origin of the difference between the spherical and the 
deformed systems is not clear. 
In order to clarify the difference in the diffuseness parameter, 
apparently 
further precision measurements for large-angle quasi-elastic scattering 
at deep sub-barrier energies are urged, especially for
deformed systems. 

\section{
Coupled-channels calculations for quasi-elastic barrier distribution
for $^{70}$Zn + $^{208}$Pb reaction}

We next discuss the barrier distribution for 
synthesis of superheavy elements. 
When one discusses a 
fusion reaction to synthesize superheavy elements, one often refers  
to a single potential such as the Bass barrier \cite{Bass}. 
On the other hand, 
the effect of channel coupling is in general strong 
for massive systems, and thus one can expect a broad distribution 
of potential barriers. 
It is thus important to study how the potential barrier is 
distributed for massive systems, since 
it is crucial to choose the right beam energy in order to 
effectively synthesize superheavy elements. 
Moreover, there is no a priori evidence why the Bass barrier is reasonable 
in the superheavy region. 
For these reasons, the quasi-elastic 
barrier distribution measurements have been recently performed by 
Mitsuoka {\it et al.} for systems relevant to 
cold fusion reactions, 
$^{48}$Ti, $^{54}$Cr, $^{56}$Fe, $^{64}$Ni, 
$^{70}$Zn + $^{208}$Pb \cite{I06}. 
In this section, we perform coupled-channels
calculations for the $^{70}$Zn + $^{208}$Pb system. 

The calculations are done with a version \cite{CQUEL} of 
the coupled-channels code {\tt CCFULL} \cite{ccfull}. 
This code treats the coupling to all orders in the coupling
hamiltonian and 
employs the isocentrifugal approximation in order
to reduce the dimension of the coupled-channels equations.
It has been shown that the isocentrifugal approximation works well 
for quasi-elastic scattering at backward angles \cite{HR04}. 
In the code, the regular boundary condition is imposed at the
origin, instead of the incoming boundary condition. 

\begin{figure}
  \includegraphics[height=.45\textheight]{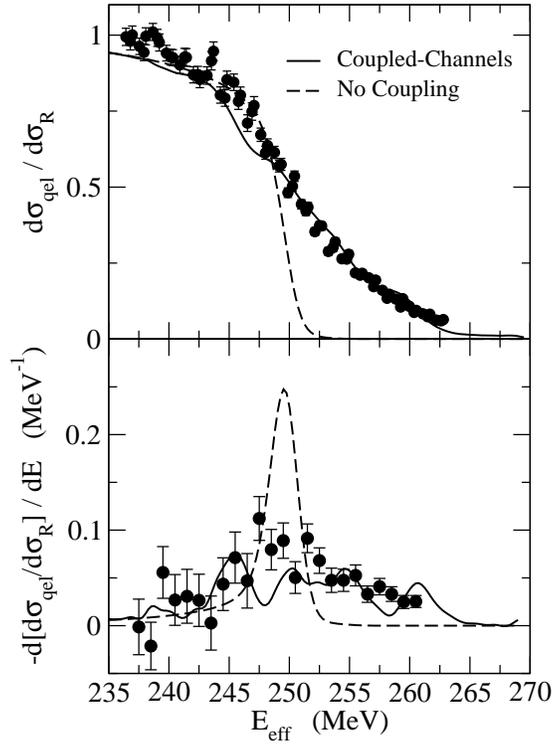}
  \caption{
The ratio of the quasi-elastic to the Rutherford cross sections 
(the upper panel) and the quasi-elastic barrier distribution (the 
lower panel) for the $^{70}$Zn + $^{208}$Pb reaction. 
These are plotted as a function of effective energy defined 
by Eq. (8). 
The solid line is the solution of coupled-channels equations, 
which take into account the double quadrupole phonon excitations in
the $^{70}$Zn nucleus and the triple octupole phonon 
excitations in the $^{208}$Pb nucleus. The dashed line shows the 
result without the couplings. 
The experimental data are taken from Ref. \cite{I06}. 
}
\end{figure}

Figure 2 shows the excitation function of the quasi-elastic 
scattering (the upper panel) and the barrier distribution (the lower
panel). 
The deep-inelastic component has been subtracted from the experimental
data using a statistical code, as is explained in Ref. \cite{I06}. 
The solid and dashed lines are the results of the
coupled-channels and the potential model calculations, respectively. 
We use the Woods-Saxon potential with $V_0$ = 140 MeV,
$r_0$ = 1.186 fm, and $a$ = 0.69 fm for the real part and 
$W_0$ = 50.0 MeV, $r_w$ = 1.0 fm, and $a_w$ = 0.1 fm for the imaginary 
part. In the coupled-channels calculation, we include the double
quadrupole phonon excitations in the $^{70}$Zn and the triple octupole 
phonon excitations in the $^{208}$Pb nucleus. 
In addition, we include the mutual
excitation channels, [1,1], [1,2], [2,1], and [2,2], where
[$n_P$,$~n_T$] denotes the excitation channel with $n_P$ phonon 
state in the projectile and $n_T$ phonon state in the target
nucleus. In this way, we include 10 channels (including the entrance
channel, [0,0]) in the calculations. 
The excitation energy for the single phonon state and the deformation 
parameter are $E_{2^+}$ = 0.885 MeV and $\beta_2$ = 0.228 for the
projectile nucleus $^{70}$Zn 
and $E_{3^-}$ = 2.614 MeV and $\beta_3$ = 0.11 for the target nucleus $^{208}$Pb. 
We use $r_0$ = 1.2 fm for the coupling term. 

In the code, the coupled-channels equations are solved by constructing
$N$ linear independent solutions of the equations, where $N$ is the
dimension of the coupled-channels equations. A linear
superposition of these solutions is then taken to 
construct the physical solution, which fulfills the asymptotic boundary
condition for scattering. For massive systems, it is
sometimes difficult to numerically maintain the linear independence of the
solutions, since the wave functions scale very differently from one
channel to another. This leads to a numerical instability of the
solution of the coupled-channels equations. We avoid this difficulty
by taking a linear superposition of the solutions at several places, 
with an interval of 1 fm up to 15 fm, so that the linear independence
is recovered. See Ref. \cite{CQUEL} for details. 
Even though we use this prescription, we still find a small 
spurious oscillation in the calculated excitation function of
quasi-elastic cross section 
due to the numerical inaccuracy, when the 
coupling is strong. We therefore average the calculated 
cross sections with a Gaussian weight with 0.5 MeV width. We have checked that
the shape of quasi-elastic barrier distribution is insensitive to the
value of the width parameter. 

As we can see in the figure, the potential model calculation (the
dashed line) significantly underestimate the quasi-elastic cross 
sections at energies above the Coulomb barrier. Also, the barrier 
distribution has a significantly narrow width, and is inconsistent with the 
experimental data. On the other hand, the coupled-channels calculation 
(the solid line) well reproduces the experimental data both for 
the excitation function and barrier distribution. 

\begin{figure}
  \includegraphics[height=.45\textheight]{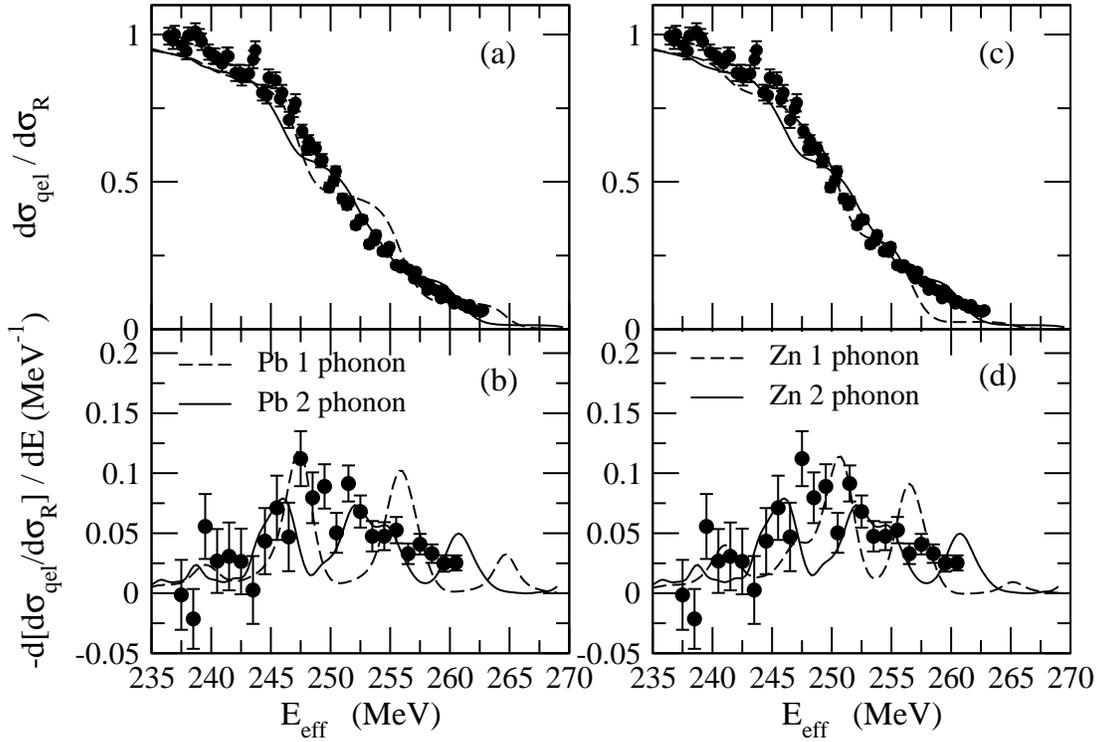}
  \caption{
The ratio of the quasi-elastic to the Rutherford cross sections 
(3(a) and 3(c)) and the quasi-elastic barrier distribution (3(b) and
3(d)) for the $^{70}$Zn + $^{208}$Pb reaction. 
The figs. 3(a) and 3(b) are obtained by including the different number of
octupole phonon excitations in the target nucleus as indicated in the
inset, together with 
the double quadrupole phonon excitations in the projectile nucleus. 
The figs. 3(c) and 3(d) are obtained by including the different number of
quadrupole phonon excitations in the projectile nucleus together with 
the double octupole phonon excitations in the target nucleus. 
}
\end{figure}

Figure 3 shows the role of multi-phonon excitations. 
Figures 3(a) and 3(b) are obtained by varying the number of octupole phonon 
excitations in the target while keeping the double phonon excitations
in the projectile nucleus. On the other hand, figs. 3(c) and 3(d) are 
obtained by 
varying the number of quadrupole phonon excitation while keeping the 
number of octupole phonon excitations in the target nucleus to be two. 
These figures show that the double phonon excitations considerably
alter the shape of barrier distribution as compared with the barrier distribution
obtained with the single phonon excitation. 
For both in the projectile and in the target nuclei, the double 
phonon excitation leads to better agreement with the experimental
data, although we find that the triple phonon excitation in the target nucleus
plays a much less important role. A similar conclusion 
has been obtained also in
Ref.\cite{RGH06}, where the role of multi-phonon excitations was
discussed for the $^{100}$Mo+$^{100}$Mo fusion reaction at energies around
the Coulomb barrier. 

\section{Summary}

We have discussed two recent developments in heavy-ion quasi-elastic
scattering at energies around the Coulomb barrier. 
We first discussed the surface property of internucleus potential. 
We have argued that the quasi-elastic scattering at deep subbarrier
energies offer a clear and almost model independent way to determine
the surface diffuseness parameter, that is, the slope of asymptotic 
exponential tail of the potential. 
The value of diffuseness parameter extracted from the $^{32,34}$S +
$^{197}$Au reactions is around 0.55 fm, and is consistent with the
double folding potential. 
On the other hand, the surface diffuseness parameter extracted from 
systems with a deformed target, that is, $^{16}$O + $^{154}$Sm,
$^{186}$W was found to be much larger (1.14 fm for the former and 0.79
fm for the latter systems). Further investigations will be required 
in order to clarify the system dependence of the surface diffuseness
parameter. In the second part, we performed the 
coupled-channels analyses for a cold fusion reaction 
$^{70}$Zn + $^{208}$Pb, where the quasi-elastic barrier distribution
was recently obtained by Mitsuoka {\it et al.}. Including the double
quadrupole phonon excitations in the projectile nucleus 
$^{70}$Zn and the triple octupole phonon excitations in the target
nucleus $^{208}$Pb in the coupled-channels calculation, we could
reproduce reasonably well both the excitation function of
quasi-elastic cross section and the shape of quasi-elastic barrier
distribution. This indicates that the coupled-channels approach still
works for the approaching phase of the reaction even in massive
systems, where many degrees of freedom may be
involved in the reaction \cite{RGH06}. It also suggests that the
deep-inelastic collision can be regarded as a post-barrier
phenomena, since the experimental quasi-elastic cross sections have
been obtained by subtracting the deep-inelastic components. 
We also discussed the role of multi-phonon excitations,
and showed that they play an important role in this system. The
coupled-channels analyses for other cold fusion reactions, 
$^{48}$Ti, $^{54}$Cr, $^{56}$Fe, $^{64}$Ni
 + $^{208}$Pb are now in progress, and we will report on them in a
 separate publication. 

\begin{theacknowledgments}
This work is based on collaborations with N. Rowley and K. Washiyama. 
We thank H. Ikezoe and S. Mitsuoka for useful discussions and for sending us the experimental
data before publication. 
This work was supported by the Grant-in-Aid for Scientific Research,
Contract No. 16740139 from the Japanese Ministry of Education,
Culture, Sports, Science, and Technology.
\end{theacknowledgments}

\end{document}